\documentclass[aps,prb,twocolumn,amsmath,amssymb,nofootinbib,superscriptaddress,floatfix]{revtex4-1}

\usepackage{amsmath}
\usepackage{amssymb}
\usepackage{amsthm}
\usepackage[pdftex]{color} 
\usepackage{graphicx}
\usepackage{dcolumn} 
\usepackage{bm} 
\usepackage{longtable}
\usepackage{ulem}   
\newcommand{\bs}[1]{{\boldsymbol{#1}}}

\begin{document}

\title{Spin Currents and Spontaneous Magnetization 
at Twin Boundaries of Noncentrosymmetric Superconductors
}

\author{Emiko Arahata}
\affiliation{Department of Basic Science, 
The University of Tokyo, 
3-8-1  Komaba, Meguro-ku, Tokyo,  
153-8902, Japan}

\author{Titus Neupert}
\affiliation{Theoretische Physik ETH-H\"onggerberg, 
CH-8093 Z\"urich, Switzerland
}
\affiliation{Condensed Matter Theory Group, 
Paul Scherrer Institute, CH-5232 Villigen PSI, Switzerland}

\author{Manfred Sigrist}%
\affiliation{Theoretische Physik ETH-H\"onggerberg, 
CH-8093 Z\"urich, Switzerland
}%

\date{\today}
\begin{abstract}
Twin boundaries are generic crystalline defects in noncentrosymmetric crystal structures. 
We study theoretically twin boundaries in time-reversal symmetric noncentrosymmetric superconductors that admit parity-mixed Cooper pairing. Twin boundaries support spin currents as a consequence of this parity mixing.
If the singlet and triplet components of the superconducting order parameter are of comparable magnitude,  
the superconducting state breaks spontaneously the bulk time-reversal symmetry locally near the twin boundary. By self-consistently evaluating the Bogoliubov-de-Gennes equations and the gap functions we find two distinct phases: First, time-reversal symmetry breaking enhances the spin currents but does not lead to charge current. A secondary phase transition then triggers a spin magnetization and a finite charge current near the twin boundary. 
\end{abstract}
\maketitle

Initiated by the discovery of superconductivity in the noncentrosymmetric
heavy Fermion compound CePt$_3$Si,
noncentrosymmetric superconductors (NCSC) have opened up new perspectives in the 
study of unconventional superconductivity.%
~\cite{PhysRevLett92.027003,
PhysRevLett92.097001,
JPhysSocJpn76.043712,
PhysRevB75.172511,
PhysRevB79.094504,
JPhysSocJpn77.083701,
PhysRevB82.104501,
PhysRevLett101.127003,
B_NON}
Due to the lack of inversion symmetry, 
these systems feature antisymmetric spin-orbit coupling that breaks completely the SU(2) spin-rotation symmetry.
As a consequence, the superconducting condensate has no definite parity and can be viewed as a superposition of even- and odd-parity (spin-singlet and spin-triplet)  Cooper pairs.~\cite{PhysRevLett87.037004,B_NON} The mixing ratio of even and odd-parity component is a
convenient tuning parameter to characterize the superconducting state of a NCSC. For example, quasi two-dimensional (2D) superconductors of this kind have a $\mathbb{Z}^{\ }_2$ topological attribute when fully gapped (symmetry class DIII in the classification of Ref.~\onlinecite{Schnyder08}), because the phases with dominant  even-parity pairing (odd-parity pairing) are topologically trivial (non-trivial). They are separated by a gap-closing topological phase transition.~\cite{Santos10}
Akin to the helical electronic edge states of a 2D $\mathbb{Z}^{\ }_2$ topological insulator,
helical edge modes in the form of Andreev bound states transport a non-conserved spin current along the boundary of a NCSC with dominant odd-parity pairing.  \cite{PhysRevB79.094504,PhysRevB82.104501,PhysRevB79.060505,PhysRevB76.012501,PhysRevB72.220504} 
Andreev bound states are a specific signature of unconventional Cooper pairing and directly manifest themselves in tunneling spectroscopy measurements. \cite{PhysRevLett101.127003}
 
Very generically, the topology of phases of matter can be probed at defects, such as boundaries, lattice dislocations or vortices in a superconducting order parameter.~\cite{PhysRevB.82.115120} In noncentrosymmetric materials, the crystal structure allows for another type of defect when two regions of space with the opposite inversion symmetry breaking face each other in a single crystal. In fact, the formation of such twin domains, similar to the domains in ferroelectrics, is rather likely in the crystal growth processes.
A first step towards understanding the superconducting state at twin boundaries of a noncentrosymmetric material has been undertaken in Ref.~\onlinecite{JPhysSocJpn77.083701}. It revealed that TRS can be spontaneously broken at the twin boundary (TB)
that can then host vortices enclosing fractional fluxes, which have been studied in the context of anomalous flux flow 
observed experimentally in some NCSC~\cite{JPhysSocJpn78.014705}.

In our study, we extend the phenomenology of the possible pairing states at the TB
by self-consistently evaluating the Bogoliubov-de-Gennes (BdG) equations and the gap functions. 
We find two distinct phases with broken TRS at the TB. In the first phase, the superposition of the two parity components turns complex. The spin currents, running along the TB for all ratios of parity mixing, are enhanced in this phase.
Yet, contrary to naive expectations, no charge current flows despite the broken TRS. This changes with a secondary transition to a further phase that features both a non-vanishing magnetization and an orbital supercurrent along the TB. 
In our analysis, we clarify the relation between the bulk topological phase transition and the phase diagram of the states near the TB, as well as the nature and spatial profile of the spin and charge supercurrents in each phase.


We use a tight-binding model describing a 2D NCSC with Rashba spin-orbit coupling that includes spin dependent nearest-neighbor pairing interactions which allow for the appearance of unconventional pairing channels.  The electrons hop on a square lattice $\Lambda$ of $L^{\ }_x\times L^{\ }_y$ sites $\bs{r}=(x,y) \in \Lambda$ that is spanned by the orthogonal unit vectors $\bs{a}^{\ }_i,\ i=x,y$. 
The corresponding Hamiltonian for the extended Hubbard model reads \cite{PhysRevLett87.037004} 
\begin{eqnarray}
 \begin{split}
 H:=&-\sum_{\bs{r}, i}\sum_s c_{{\bs r}+{\bs a}_i s}^\dagger
 (t \hat{\sigma}^0-\bs{\lambda}_{\bs{r},{\bs a}_i}\cdot 
 \hat{\bs{\sigma}})_{s,s^\prime}
  c_{{\bs r}s'} \\
 &+\sum_{\bs{r},i}\bigl[J{\bs S}_{{\bs r}+{\bs a}_i}\cdot {\bs S}_{{\bs r}}
 +{\bs D}_{\bs{r},\bs{a}_i}\cdot \left({\bs S}_{{\bs r}+{\bs a}_i}\times{\bs S}_{{\bs r}} \right)\bigr]
  \\
 & +U\sum_{\bs{r}}n_{{\bs r}\uparrow}n_{{\bs r}\downarrow}+V\sum_{\bs{r},i}n_{{\bs r}}n_{{\bs r}+{\bs a}_i},
 \end{split}
 \label{Hami_0}
 \end{eqnarray}
where  $c_{{\bs r} s}^\dagger$ creates an electron with spin $s=(\uparrow, \downarrow)$ at site $\bs{r}\in \Lambda$.
The antisymmetric spin-orbit coupling is a Rashba term of strength $\alpha_{\bs{r}}$ parametrized by 
${\bs \lambda}_{\bs{r},\bs{a}_i}=\mathrm{i}\alpha_{\bs{r}} (\hat{\bs{z}} \times \bs{a}_i )$, $i=x,y$, where $\hat{\bs{z}}$ is the unit vector normal to the plane of the lattice. The vector $\mbox{\boldmath  $\hat \sigma$}=(\sigma^x,\sigma^y,\sigma^z)$ denotes the three Pauli matrices and $\hat{\sigma}^0$ the $2\times2$ unit matrix. 
We define the electron density operator $n_{{\bs r}_i}=n_{{\bs r}_i\uparrow}+n_{{\bs r}_i\downarrow}$ and the spin density operator
${\bs S}_{{\bs r}}=\sum_{s,s'}c^\dagger_{\bs{r} s} \hat{\bs{\sigma}}^{\ }_{s,s^\prime}c^{\ }_{\bs{r} s'} $.
Besides the ordinary spin-isotropic Heisenberg exchange of strength $J$,
the noncentrosymmetric crystal structure also allows for a Dzyalonshinsky-Moriya type spin-spin interaction of strength $D_{\bs{r}}$
which is parametrized as $\bs{D}_{\bs{r},\bs a_i}=D_{\bs{r}}(\hat{\bs{z}} \times \bs{a}_i )$, $i=x,y$. 

Let us illustrate the mean-field decoupling for a translational invariant system ($\alpha_{\bs{r}}\equiv\alpha$, $D_{\bs{r}}\equiv D$, $\forall \bs{r}$), assuming periodic boundary conditions in both the $\bs{a}^{\ }_x$ and $\bs{a}^{\ }_y$ directions.
The Rashba-type spin-orbit interaction breaks the SU(2) spin-rotation symmetry and induces a splitting of the electron
bands, with each band exhibiting a specific spin structure in momentum space. 
At the same time, it allows the gap function to be of mixed parity. The
BdG mean-field Hamiltonian reads 
$H^{\mathrm{BdG}}:=
\sum_{\bs {k}} {\bs c}^\dagger_{\bs k}
\mathcal{H}^{\mathrm{BdG}}_{\bs{k}}
{\bs c}_{\bs k}$,
with the four-component notation ${\bs c}_{\bs k}=(c_{{\bs k}\uparrow},c_{{\bs k}\downarrow},c^\dagger_{-{\bs k}\uparrow},c_{-{\bs k}\downarrow}^\dagger)$, ${\bs k}$ representing the 2D momentum in the Brillouin zone and 

\begin{subequations}
\begin{eqnarray}
\mathcal{H}^{\mathrm{BdG}}_{\bs{k}}=
\left(\begin{array}{cc}
\mathcal{H}_{\mathrm{kin},\bs{k}} & \Delta_{\bs{k}} \\\Delta^\dagger_{\bs{k}} & -\mathcal{H}_{\mathrm{kin},-\bs{k}}^{\mathsf{T}}\end{array}\right). 
\label{H_k}
\end{eqnarray}
The kinetic part of the Hamiltonian is given by
\begin{equation}
\mathcal{H}_{\mathrm{kin},\bs{k}}:=
-2t[\cos(k_x)+\cos(k_y)]
\hat{\sigma}^0+{\bs g}_{\bs k}\cdot \hat{\bs{\sigma}},
\end{equation}
with
${\bs g}_{\bs{k}}:=2\alpha[\bs{a}^{\ }_x \sin(k_y)-\bs{a}^{\ }_y \sin(k_x)]\nonumber$.
We can decompose the superconducting gap function
$\Delta_{\bs k}=(\mathrm{i}\sigma^{y})\bigl(\Delta^{(\mathrm{e})}_{\bs k}+\bs{\Delta}^{(\mathrm{o})}_{\bs k}\cdot\hat{\bs{\sigma}}\bigr)$
in a scalar even-parity spin singlet part $\Delta^{(\mathrm{e})}_{\bs k}$ and a vector odd-parity spin-triplet part $\bs{\Delta}^{(\mathrm{o})}_{\bs k}$. \cite{B_NON,RevModPhys63.239}
The assumption of an extended $s$-wave pairing in the former and $p$-wave pairing in the latter yields the momentum dependences 
\begin{eqnarray}
\Delta^{(\mathrm{e})}_{\bs k}&=&\Delta_s^{(\mathrm{e})}\left[ \cos(k_x)+ \cos(k_y)\right]+\Delta_0^{(\mathrm{e})}, 
\\
\bs{\Delta}^{(\mathrm{o})}_{\bs k}&=&\Delta_p\left( \sin (k_y),-\sin (k_x) ,0\right).
 \end{eqnarray}
of the order parameters. Note that the $p$-wave pairing state yielding the highest transition temperature is the one with $\bs{\Delta}^{(\mathrm{o})}_{\bs k}\propto \bs{g}_{\bs{k}}$.~\cite{B_NON}
 \end{subequations}
This allows to simultaneously diagonalize the gap function and the kinetic part of the Hamiltonian $\mathcal{H}_{\mathrm{kin},\bs{k}}$ by going to the basis of $\lambda=\pm$ helicity states that label the two Fermi surface sheets. 
The gap function on either sheet can be conveniently represented in the mixed-parity parametrization, 
$\Delta_{\lambda,\bs{k}}=\Delta^{(\mathrm{e})}_{\bs{k}}+\lambda\bs{\Delta}^{(o)}_{\bs{k}}\cdot\bs{g}_{\bs{k}}/|\bs{g}_{\bs{k}}|$, $\lambda=\pm$.

\begin{figure}[t]
\includegraphics[width=86mm]{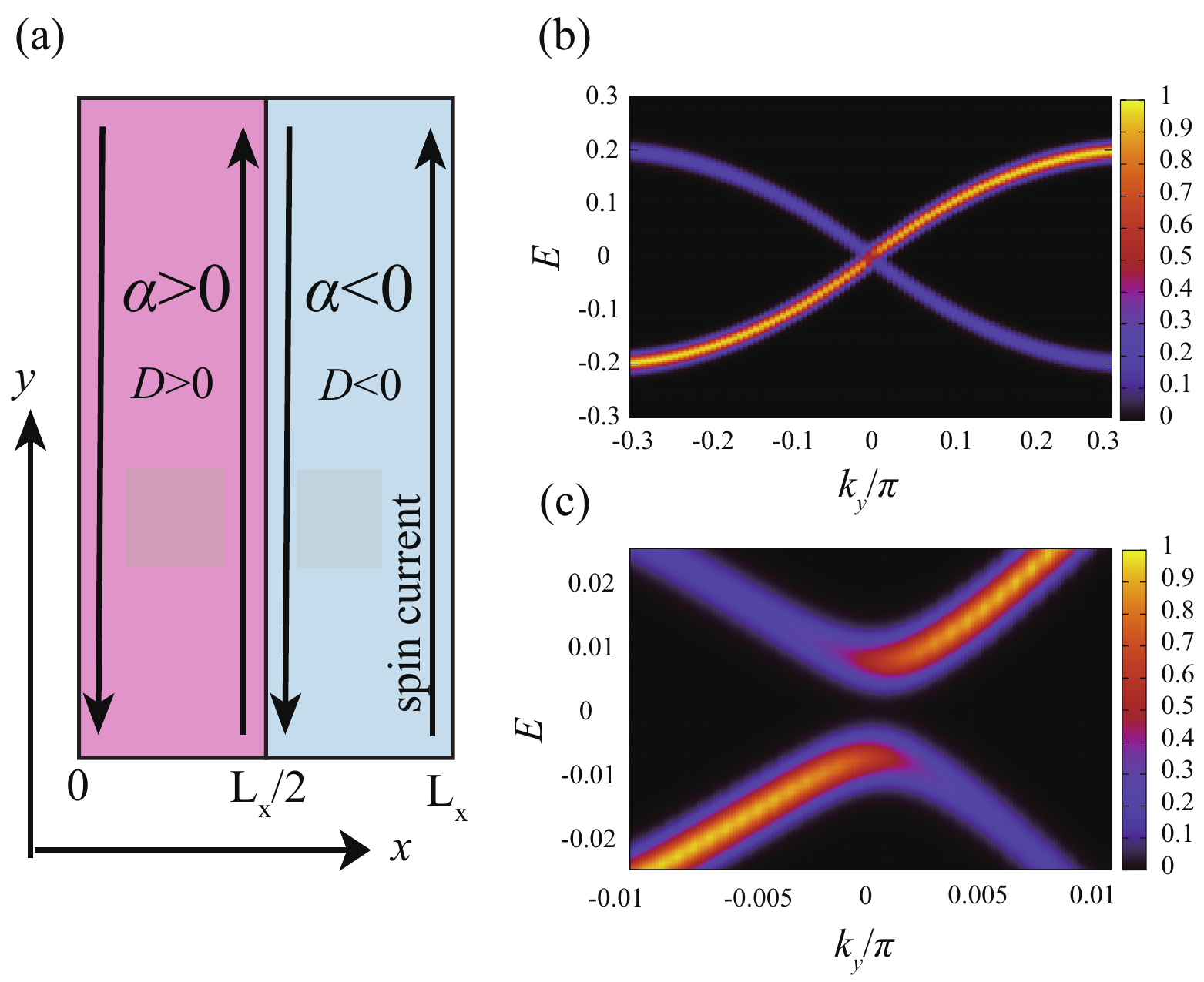}
\caption{(Color online) (a) Sketch of the intrinsic metallic interface in a NCSC.
If the pairing is dominantly of triplet type, a topologically protected Kramers pair of edge modes carrying a spin current is localized at $x=0$ and $x=L_x$.
Two Kramers pairs of edge modes that are localized near the TB hybridize and thereby loose their topological protection. 
 (b) The spectral function of up-spin quasiparticles at the immediate left of the TB , $A^{\uparrow}_{x=50,k_y}(E)$ shows the spin-polarization of these helical states. The zoom-in (c) reveals their hybridization. ($A^{\downarrow}_{x=50,k_y}(E)$ and $A^{\uparrow}_{x=51,k_y}(E)$ are obtained by flipping the figure about $k_y=0$.) Parameters are $J=1.3$, $D=1.75$, $V=1.22$, $U=0.82$, such that $\Delta_s/\Delta_p\sim0.425$.~\cite{NoteParameters}
 }
\label{Fig:energy}
\end{figure}

We now turn to the electronic properties of the system with TB. For that, we equip the Rashba spin-orbit coupling and the Dzyalonshinsky-Moriya interaction
in Hamiltonian~\eqref{Hami_0} with the spatial dependencies
\begin{equation}
\alpha^{\ }_{\bs{r}}=\alpha\, \mathrm{sgn}(x-L_x/2),
\qquad
D^{\ }_{\bs{r}}=D\, \mathrm{sgn}(x-L_x/2),
\end{equation}
respectively.
This models a TB located at $x=L_x/2$ and separates two regions of the superconductor that have the opposite sign of the Rashba and Dzyalonshinsky-Moriya coupling [see Fig.~\ref{Fig:energy} (a)].
Open and periodic boundary conditions are used in the $\bs{a}^{\ }_x$ and $\bs{a}^{\ }_y$ directions, respectively.
The relative U(1) phase between $\Delta^{(\mathrm{e})}$ and $\Delta_p$ has to change by $\pi$ across the TB. The way the superconducting condensate accommodates this phase twist decisively determines the physics at the TB.

If $\bs{\Delta}^{(\mathrm{o})}$ is the dominant component of the order parameter, gapless helical edge states exist at the boundary of a 2D topological superconductor, as dictated by the nontrivial $\mathbb{Z}_2$ topological index.
Localized modes within the bulk spectral gap also exist at the TB, though they would not be endowed with topological protection, for the $\mathbb{Z}_2$ topological sector of the Hamiltonian is the same on either side.
In order to identify the existence and spin polarization of the localized modes at TB, we calculated the spectral function 
\begin{eqnarray}
 A^{s}_{x,k_y}(E)&=&-\frac{1}{\pi}{\rm Im} \; G^{s,s}_{x,k_y}(E),
   \end{eqnarray}
where $G^{s,s}_{x,k_y}$ is the Green's function at position $x$ and momentum $k_y$.
Figures~\ref{Fig:energy}(b) and~(c) show the spectral function for up-spin quasiparticles on the immediate left of the TB. We observe that the left- and rightgoing modes have opposite spin polarization at high energies and that this spin-momentum locking is lifted with the appearance of a hybridization gap near zero energy.  

\begin{figure}[t]
\includegraphics[width=85mm]{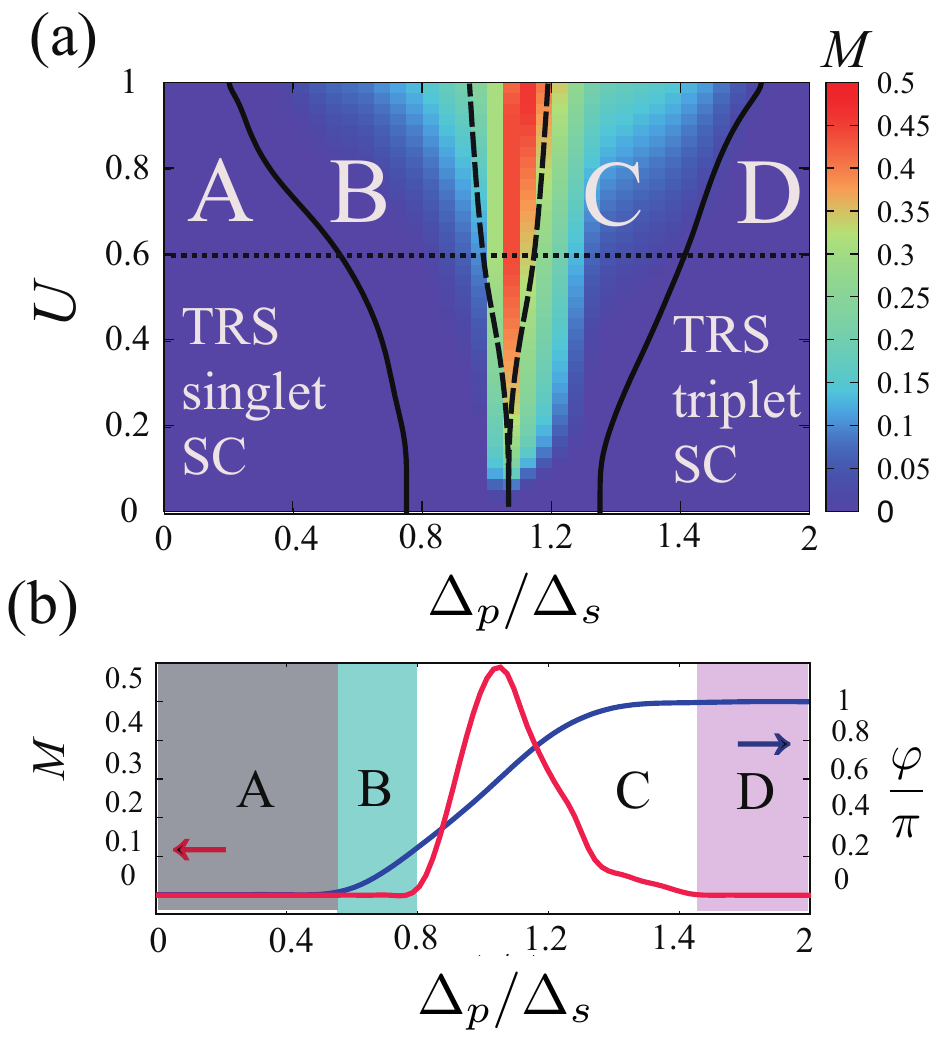}
\caption{(Color online) (a) Phase diagram of superconducting and magnetic order at the TB as a function of on-site repulsion $U$ and $D=1.8$.~\cite{NoteParameters} The parameters $J$ and $V$ are varied to tune the mixing ratio.
Continuos lines separate phases of 
(A) TRS singlet dominated superconductivity, 
(B,C) TRS breaking phase where the $\varphi$ deviates from the values 0 and $\pi$ near the TB, and
(D) TRS triplet dominated superconductivity. The color scale measures the spin magnetization near the TB that becomes nonzero in phase (C) via a secondary phase transition in the phase with broken TRS.
Dashed lines encircle the region in parameter space where the bulk superconducting gap is nodal, marking the topological phase transition from the trivial (left) to the nontrivial (right) $\mathbb{Z}^{\ }_2$ topological superconductor. 
(b) Spin-magnetization $M$ and phase $\varphi$ along the dotted line in (a).
}
\label{Fig:diagram}
\end{figure}

We turn now to the central issue, the analysis of TRS breaking phase at the TB. TRS breaking is signaled by the following two quantities: (i) the relative U(1) phase 
\begin{equation}
\varphi:=\mathrm{arg}\Delta^{(\mathrm{e})}_{x=L_x/4}-\mathrm{arg}\Delta^{(\mathrm{e})}_{x=3L_x/4}
\end{equation}
of the singlet component of the superconducting order parameter in the bulk on the left and right side of the TB and
(ii) the spin magnetization  $M\propto n_{\uparrow}-n_{\downarrow}$ at the TB.
To understand the relevance of (i), we have to ask how the condensate can account for the $\pi$-shift in the relative U(1) phase between $\Delta^{(\mathrm{e})}$ and $\Delta_p$ across the TB.
The two values of $\varphi$ compatible with TRS are $\varphi=0$ and $\varphi=\pi$. In the former case for $|\Delta^{(\mathrm{e})}|\gg |\bs{\Delta}^{(\mathrm{o})}|$ and it is energetically favorable that $\Delta_p$ changes the sign being zero at the TB. Conversely, if $|\Delta^{(\mathrm{e})}|\ll |\bs{\Delta}^{(\mathrm{o})}|$,  the sign change should occur on the $\Delta^{(\mathrm{e})}$ component such that $\varphi=\pi$. 
If $|\Delta^{(\mathrm{e})}|$ and $|\bs{\Delta}^{(\mathrm{o})}|$ are of comparable magnitude, the system is frustrated, since the cost in condensation energy for a node  at the TB in either component is high. In this case, the relative U(1) phase moves continuously from $0$ to $\pi$ across the TB, thereby breaking TRS. This is also signaled by $\varphi\neq0,\pi$, where either of the degenerate solutions $\varphi$ and $-\varphi$ is spontaneously chosen.
The phase diagram in Fig.~\ref{Fig:diagram} shows that the local TRS breaking at the TB as identified through $\varphi$ [continuous lines in Fig.~\ref{Fig:diagram}(a)], covering region B and C, and the finite magnetization $M$ (color code), restricted to region C, occur within two different phase boundaries. The secondary phase transition towards the TRS breaking phase C with finite magnetization $M$ arises inside the TRS breaking phase with $\varphi\neq0,\pi$. 

For comparison, the dashed lines in Fig.~\ref{Fig:diagram}(a) define the region in parameter space where the bulk system is gapless due to nodes in the superconducting order parameter. Left (right) of this region, the bulk is a topologically trivial (non-trivial) superconductor. The topological phase transition acquired a finite width due to the term $\Delta^{(\mathrm{e})}_s$ in the Hamiltonian. Inside the gapless region, the order parameter has several  point nodes on the Fermi surfaces. In the edge Brillouin zone, flat edge bands stretch between the projections of pairs of these nodal points, much like the flat bands at zigzag edges of grapheme (see also Ref.~\onlinecite{PhysRevB.84.060504}).~\cite{JPSJ65} 
We make the following two observations that relate to these topological features. First, the bulk topological phase transition happens fully inside the parameter range in which TRS is spontaneously broken at the TB. Second, the magnetic order at the TB is strongly enhanced in the bulk gapless region. This can be attributed to Stoner-like magnetism of the flat bands at the TB.~\cite{JPSJ65}  

Finally, let us study how the spontaneous TRS breaking manifests itself in the context of a (spin) Hall response at the TB.
Spin currents are generically expected at edges and TB in a noncentrosymmetric materials.~\cite{PhysRevB.78.153302,PhysRevLett.106.237201}
\begin{figure}[t]
\includegraphics[width=85mm]{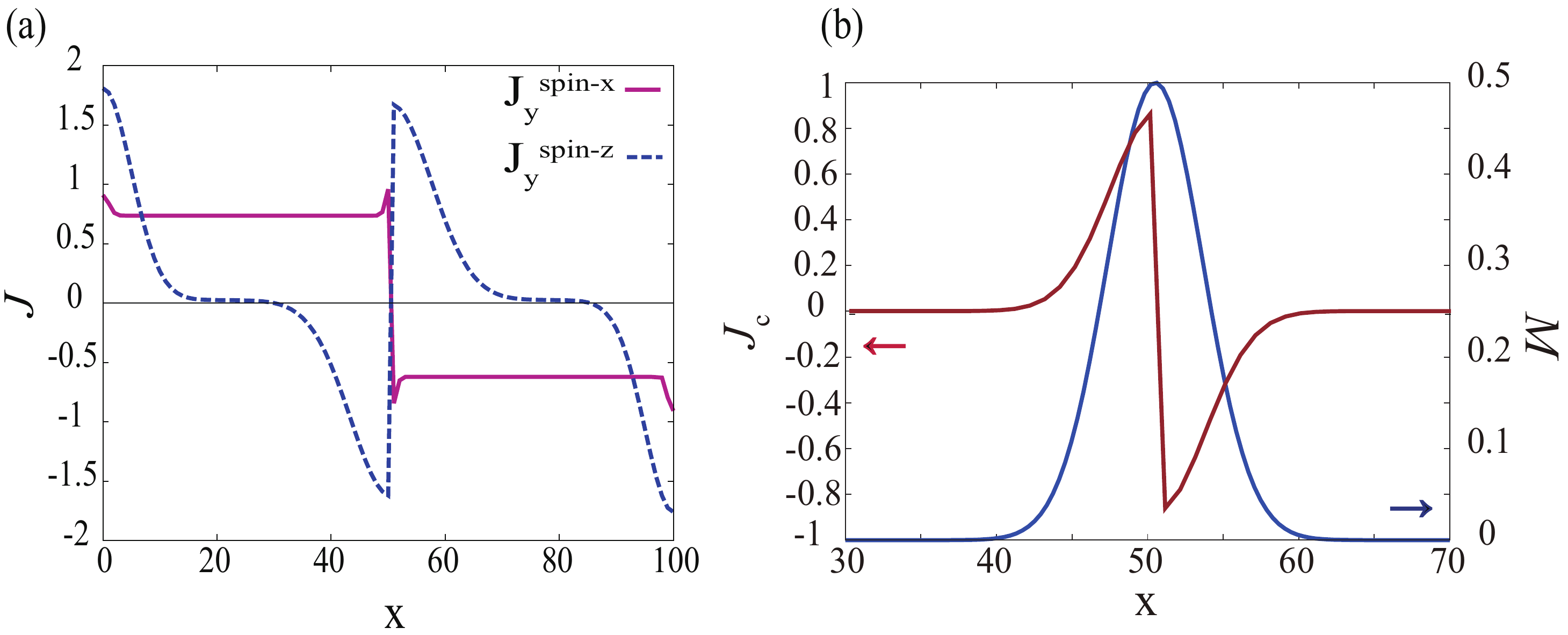}
\caption{(Color online) (a) Spin current in the $y$ direction as a function of $x$. The solid lines and the dashed lines are the spin currents with spin polarization in the $x$ and $y$ direction, respectively. 
Parameters are $J=1.1$, $D=2.0$, $V=1.5$, $U=0.87$, such that $\Delta_s/\Delta_p=0.41$.~\cite{NoteParameters}
(b) The $x$-dependence of 
the orbital supercurrent (red line) and of the magnetization (blue line) near the TB in the TRS broken phase.
Parameters are $J=1.3$, $D=2.1$, $V=1.2$, $U=0.87$, such that $\Delta_s/\Delta_p=0.5$.~\cite{NoteParameters}
}
\label{Fig:spincurrent}
\end{figure}
We define the spin current of polarization $i=x,y,z$ that runs in the $y$-direction 
   \begin{eqnarray}
J_y^{{\rm spin-}i}(x):=
{\rm Tr} \sum_{ k_y} \bs{c}^\dagger_{k_y,x}\left(\partial_{k_y} \mathcal{H}^{\mathrm{BdG}}_{k_y,x} \right)\hat \sigma^i \bs{c}_{k_y,x},
\label{spincu}
   \end{eqnarray}
where the trace is taken over all states below zero energy.  
Figure~\ref{Fig:spincurrent}(a) shows the spin currents $J_y^{{\rm spin-}x}$ and $J_y^{{\rm spin-}z}$ as a function of position $x$ inside the phase B of phase diagram~\ref{Fig:diagram}(a). Each of them has opposite signs on either side of the TB. 
The component $J_y^{{\rm spin-}y}$ vanishes in the bulk and is much smaller than $J_y^{{\rm spin-}z}$ at the TB (not shown).
The component $J_y^{{\rm spin-}z}$ corresponds to the usual spin current, present at the sample edge as well as at the TB. 
The component $J_y^{{\rm spin-}x}$ is finite also in the bulk and increases at the boundary. However, its bulk contribution should not be interpreted as a physically measurable spin current.~\cite{PhysRevB.68.241315}
The phase C of phase diagram~\ref{Fig:diagram}(a) is characterized by a finite magnetization $M$ shown in Fig.~\ref{Fig:spincurrent}(b). This magnetization appears together with a orbital supercurrent that runs in opposite directions on the immediate left and right of the TB [Fig.~\ref{Fig:spincurrent}(b)]. This may be considered a spontaneous spin Hall effect as the supercurrent is a response to introducing an imbalance of the spin occupation on the spin current.  

In summary, we studied the interface states between twin domains in a mixed-parity superconductor near a topological phase transition connecting a topologically trivial with a non-trivial phase. A sequence of two phases localized around the twin boundary appear. The primary phase breaks TRS and is, consequently, two-fold degenerate, but does not show any magnetism. The secondary phase introduces magnetism through spin polarization along z-axis and a supercurrent parallel to the TB. This phase breaks the reflection symmetry for a mirror plane perpendicular to the TB including the z-axis, adding a further two-fold degeneracy. The appearance of the spin polarization and the supercurrent are connected through the presence of spin current at the TB analogous to the spin Hall effect.~\cite{footnote:Schnyerpaper} 

In closing, we note that the mechanism for TRS breaking that we discussed in this work is not limited to twin boundaries in NCSCs. It can also be of relevance to tunable devices with interface superconductivity such as SrTiO$_3$/LaAlO$_3$, if the Rashba spin-orbit coupling is not uniform in space.~\cite{Reyren07} A similar phenomenology might apply to other ordering phenomena, such as the spontaneous generation of the quantum spin Hall effect in graphene-like materials.~\cite{Raghu08} In this case, TRS may be broken at the boundary between regions of opposite spin-Hall conductivity, spontaneously generating a charge Hall effect.   

We would like to thank M. Achermann, A. Schnyder and P. Brydon for helpful discussions.  
This work was supported in part by the Swiss National Science Foundation, the NCCR MaNEP and Pauli Center 
for Theoretical Studies of the ETH Zurich. 
E. A. was supported by a Grant-in-Aid from JSPS.

\bibliographystyle{apsrev}
\bibliography{Double}

\begin{thebibliography}{27}
\bibitem[{Not()}]{NoteParameters}
\bibinfo{howpublished}{Throughout this paper, we set $\alpha=1$, $L_x=100$ and
  fix the electron density per lattice site to $n=0.9$, while we change the
  parameters of the interaction $U$, $V$, $J$, and $D$ as appropriate. All
  energies are measured in units of $t$.}
\end{thebibliography}
\end{document}